\documentclass[sn-mathphys,Numbered]{sn-jnl}% Math and Physical Sciences Reference Style
\usepackage{graphicx}%
\usepackage{multicol}% added by Julia
\usepackage{multirow}%
\usepackage{amsmath,amssymb,amsfonts}%
\usepackage{amsthm}%
\usepackage{mathrsfs}%
\usepackage[title]{appendix}%
\usepackage{xcolor}%
\usepackage{textcomp}%
\usepackage{manyfoot}%
\usepackage{booktabs}%
\usepackage{algorithm}%
\usepackage{algorithmicx}%
\usepackage{algpseudocode}%
\usepackage{listings}%
\usepackage{float}% added by Julia
\usepackage{array}%added by Julia
\newcolumntype{L}[1]{>{\raggedright\let\newline\\%added by Julia
\arraybackslash\hspace{0pt}}m{#1}}%added by Julia
\newcolumntype{C}[1]{>{\centering\let\newline\\%added by Julia
\arraybackslash\hspace{0pt}}m{#1}}%added by Julia
\newcolumntype{R}[1]{>{\raggedleft\let\newline\\%added by Julia
\arraybackslash\hspace{0pt}}m{#1}}%added by Julia
\theoremstyle{thmstyleone}%
\theoremstyle{thmstyletwo}%

\theoremstyle{thmstylethree}%

\begin{document}

\title[Parameter uncertainty estimation for exponential semi-variogram models: Two generalized bootstrap methods with check- and quantile-based filtering]{Parameter uncertainty estimation for exponential semi-variogram models: Two generalized bootstrap methods with check- and quantile-based filtering}


\author*[1]{\fnm{Julia} \sur{Dyck}}\email{j.dyck@uni-bielefeld.de}

\author[1,2]{\fnm{Odile} \sur{Sauzet}}\email{odile.sauzet@uni-bielefeld.de}

\affil*[1]{\orgdiv{Department of Business Administration and Economics}, \orgname{Bielefeld University}, \orgaddress{\street{Universitätsstraße 25}, \city{Bielefeld}, \postcode{33615}, \country{Germany}}}

\affil[2]{\orgdiv{School of Public Health}, \orgname{Bielefeld University}, \orgaddress{\street{Universitätsstraße 25}, \city{Bielefeld}, \postcode{33615}, \country{Germany}}}


\abstract{Estimation of parameter standard errors for semi-variogram models is challenging, given the two-step process required to fit  a parametric model to spatially correlated data. 

Motivated by an application in social epidemiology, we focus on exponential semi-variogram models fitted to data with 500 to 2000 observations and little control over the sampling design. Previously proposed methods for  the estimation of standard errors cannot be applied in this context. Approximate closed form solutions are too costly using generalized least squares in terms of memory capacities. The generalized bootstrap proposed by Olea and Pardo-Ig\'{u}zquiza is nonetheless  applicable with weighted instead of generalized least squares. However, the standard error estimates are hugely biased and imprecise. 

Therefore, we propose to add a filter mechanism to the generalized bootstrap. The new development is presented and evaluated with a simulation study which shows that the generalized bootstrap with check-based filtering leads to massively improved results compared to the generalized bootstrap method without filtering. We provide a case study using birthweight data.}




\maketitle

\section{Introduction}

\subsection{Contextual background}
Semi-variogram modelling is a method used in spatial statistics to model the relationship between pairwise correlation of observations and the distance between these observations. It is commonly used in fields like agriculture or ecology but has also seen applications in social epidemiology to study neighbourhood effects on health. 
In the latter, the estimation of exponential parametric models is of particular interest because it provides a measure of which part of the total variance is spatially structured and which local spatial scale is relevant. However, the assessment of parameter uncertainty presents some difficulties while being necessary for proper inference \cite{bard1974}.

Contrary to some other fields of application, data in social epidemiology cannot be sampled according to a scheme designed to facilitate the estimation of parameters. On the contrary, analyses are usually based on survey data in which there are few observations at very small distances \cite{sauzet2021} - providing very few data for the estimation of the first points of the empirical semi-variogram - but with possibly large sample sizes  of over 1000 -  making, for instance, inverting matrices for the estimation of standard errors difficult. Exemplary applications of semi-variogram modelling in this context can be found in \cite{sauzet2021} and \cite{breckenkamp2021method}.

%The technical challenges linked to the estimation of parameters associated to correlated data [add ref], the estimation of uncertainty for semi-variogram model parameters, as well, requires taking the correlation structure into account for the uncertainty estimator construction. 

\subsection{Technical challenges}

Three characteristics of semi-variogram estimation are a source of technical difficulties for the construction of parameter estimators and, thereby, for the construction of parameter uncertainty estimators.
First, the estimation of parameters associated with correlated data has to take the correlation structure into account. Consequently, the simplifying assumption of an underlying data generating process with independent and identically distributed elements is not appropriate \cite[p.3]{Cressie.1993}. Second, the estimation of the semi-variogram is not based on the spatial observations themselves but on their pairwise Euclidean distances \cite{Webster.2007}. Third, the fitting of a semi-variogram model requires a two-step process that includes, first, the estimation of a discrete semi-variogram and, second, based on that the fitting of a mostly continuous model. Multiple steps of transformation and estimation intuitively lead to multiple potential sources of uncertainty that influence the precision of the semi-variogram model parameters. 
\\

\subsection{Existing approaches}
Methods for the parameter standard error estimation have been proposed in the past. 

An analytic solution was developed by Pardo-Igúzquiza and Dowd \cite{Pardo2001} for Gaussian data generating processes. Parameter estimates are obtained using the generalized least squares (GLS) estimation approach.
While it leads to reliable standard error estimates for datasets with up to 100 observations, results are often not obtained for larger sample sizes of 1000 or higher due to the at least quadratically increasing amount of memory capacities required for the computation. %Even substituting the GLS approach with the computationally less costly weighted least squares (WLS) estimation approach does not sufficiently compensate for the numerical costs. 
Moreover, the theoretical statistical assumptions made in order to construct the analytic solution are often not fulfilled such that solutions, if feasible, have to be interpreted to be approximate rather than exact.

An empirical approach using the so called generalized bootstrap is provided by Olea and Pardo-Igúzquiza \cite{olea2011generalized}.
The reduced amount of theoretical assumptions and the use of the bootstrap approach which can be implemented quite efficiently are advantageous. However, as for the previous method, the generalized bootstrap is originally constructed with focus on data with less than 100 observations, high control over the sampling design and locations, and parameter estimates obtained with the GLS approach as is typically the case in the geological context. Larger datasets and the often conducted parameter estimation based on bootstrap samples have forced us to substitute the GLS parameter estimation by the weighted least squares (WLS) estimation procedure in order to make it computationally feasible with our data. Given these changes, the standard error estimates are highly biased upwards and very imprecise.%\textcolor{red}{\footnote{master thesis simulation study}}

A more recent attempt by Saha and Datta \cite{saha2018brisc} uses an approximate maximum likelihood estimation algorithm with sparse distance matrices for the parameter estimation to fasten the computation procedure of the standard error estimates via generalized bootstrap. The approach focuses on applications to very large datasets with 10000 or more data points and assumes approximate Gaussianity for the maximum likelihood estimation. 
Datasets with a sample size of about 1000 appear to be too small to account for the errors due to sparse computation, so that the semi-variogram parameter estimators themselves are hugely biased 
%(see Tables \ref{table_par_ests_nugget}, \ref{table_par_ests_sigma}, \ref{table_par_ests_phi} in appendix B) %\textcolor{red}{\footnote{side task in performed simulation study, results can be shown in additional material}} 
and therefore not an appropriate base for uncertainty estimation.\\

\subsection{Extension of the generalized bootstrap approach}\label{sec:motivation_filter}

A difficulty associated with the used of bootstrap methods to estimate the uncertainty of parameter estimates is that the fitting of a parametric model to an empirical semi-variogram requires tuning of metaparameters and a visual check of the fit. The latter cannot be performed by a bootstrap and may lead to an overestimation of the the variability of estimates.
It follows that estimating standard errors based on one of the existing methods does not provide satisfactory results in the data scenarios of our interest.

Focusing on typical  social epidemiological data to which exponential semi-variogram models are to be fitted, we need a standard error estimation tool that
\begin{enumerate}
	\item works under as few distributional assumptions as possible,
	\item is based on the commonly used and computationally feasible WLS model fitting approach,
	\item is applicable to data with sample sizes of about 500 to 2000 and little or no control over the sampling design,
	\item is less biased and more stable that the methods available.
\end{enumerate}

We propose to use the generalized bootstrap by Olea and Pardo-Ig\'{u}zquiza \cite{olea2011generalized} using WLS rather than GLS for standard error estimation. To compensate for the higher variance of the standard error estimator 
 that is observed in simulations representing the scenarios of our interest, the generalized bootstrap procedure is extended by a filter. The filter evaluates each bootstrap estimate and excludes results of a bootstrap estimation if the semi-variogram model fitting algorithm is suspected to have not converged or provides an estimate which is a poor fit of the empirical semi-variogram.
Two filter options are presented and evaluated: the check-based filtering and the quantile-based filtering. 

\section{Methods}

\subsection{Statistical model}
We assume an isotropic and second-order stationary spatial process $\{Z(\mathbf{s}) \mid \mathbf{s} \in D\}$ with a collection of random variables $Z(\mathbf{s})$ and a set of locations $D$.
In this case, the semi-variogram of a spatial process characterizes the semi-variance structure of the process depending on the Euclidean distance $h = || \mathbf{s} - \mathbf{s}'||$ between two subjects \cite[p.135--140]{Schabenberger}. The semi-variogram is defined as 
\begin{align}
\gamma(h) = \frac{1}{2} Var(Z(\mathbf{s})- Z(\mathbf{s}'))
\end{align}
for two subjects at locations $\mathbf{s}$ and $\mathbf{s}'$ with distance $h := ||\mathbf{s} - \mathbf{s}'||$.

Under the stated conditions the semi-variogram and the covariance function of the spatial process have the relationship:
\begin{align}
C(h) = c_0 + \sigma_0^2 - \gamma(h)
\end{align}
where $c_0 + \sigma_0^2 = Var[Z(\mathbf{s})]$ is the variance of the outcome to be analysed. It is assumed to be constant over the studied surface. The summand $c_0$ is called nugget effect and defines the value of $\gamma(h)$ for $h$ approaching zero. The so called partial sill $\sigma_0^2$ describes the asymptotic covariance of two observations on the same location.

Given data, Matheron's estimator provides an unbiased empirical estimate of the semi-variogram \cite{Matheron1963}. It is defined as
\begin{align}
\hat{\gamma}(h) = \frac{1}{2\cdot|N(h)|} \sum_{(\mathbf{s_i}, \mathbf{s_j}) \in N(h)}\{Z(\mathbf{s_i})- Z(\mathbf{s_j})\}^2,
\end{align}
where $N(h) := \{(\mathbf{s_i}, \mathbf{s_j}) \in D |\;\; ||\mathbf{s_i}-\mathbf{s_j}|| = h\}$. To guarantee a sufficient cardinality of $N(h)$ for each realized $h$, multiple lag distances are grouped into lag intervals of the form $[h- \varepsilon, h + \varepsilon]$ for a discrete number of lags $h_1,\dots, h_K$. 

Whereas the empirical semi-variogram provides point estimates for a finite number of lag intervals, an almost continuous model is obtained by a parametric semi-variogram model.
The exponential semi-variogram model is defined as
\begin{align}
\gamma_{\text{exp}}(h; c_0, \sigma^2_0, \phi) = 
c_0 + \sigma_0^2 \Big\{1 - \exp\big(- \frac{h}{\phi}\big)\Big\} \;\; & \text{for}\;\; h > 0
\end{align}
and $\gamma_{\text{exp}}(h; c_0, \sigma^2_0, \phi) = 0$ for $h = 0$ with nugget effect $c_0>0$, partial sill $\sigma_0^2>0$ and range parameter $\phi>0$ \cite[p.61]{Cressie.1993}.

\subsection{Model fitting}\label{modelfitting}
The semi-variogram model fitting is commonly done via the method of WLS or GLS \cite[p.163--166]{Schabenberger}.
In both cases the parameter vector $\theta = (c_0, \sigma_0^2, \phi)^t$ that minimizes the loss function
\begin{align}\label{min_term}
	LOSS(\theta) = 
[\hat{\mathbf{\gamma}}-\mathbf{\gamma_{\text{exp}}(\theta)}]^t\; \mathbf{\Gamma}^{-1} \; [\hat{\mathbf{\gamma}}-\mathbf{\gamma_{\text{exp}}(\theta)}]
\end{align}
is chosen. The vector $\hat{\mathbf{\gamma}}= (\hat{\gamma}(h_1), \dots, \hat{\gamma}(h_K))^t$ contains the empirical semi-variogram evaluated at $K$ lag distances and $\mathbf{\gamma_{\text{exp}}(\theta)} = (\gamma_{\text{exp}}(h_1;\mathbf{\theta}), \dots, \gamma_{\text{exp}}(h_K; \mathbf{\theta}))^t$ is the corresponding semi-variogram model evaluated at the same lag distances. 
To fit a model $\gamma_{\text{exp}}$ to the empirical semi-variogram $\hat{\gamma}$ an estimate for the covariance matrix of the empirical semi-variogram denoted by $\mathbf{\Gamma} \in \mathbb{R}^{K\times K}$ is needed.
%In case of the LS method, $\mathbf{\Gamma} = \sigma^2 \mathbf{I}$ is assumed meaning that the variance for each lag's estimate is constant the lags' semi-variances are uncorrelated.
%The loss function then reduces to:
%\begin{align}\label{min_term}
%	LOSS_{LS}(\theta) = 
%	\sum_{k=1}^K [\hat{\gamma(d_k)}-\gamma_{\text{exp}}(d_k;\theta)]^2
%\end{align}

In case of the GLS estimation method, varying variances of the lags' semi-variogram estimates are allowed, as well as covariances between two lags. This is reflected by a $\mathbf{\Gamma}$ matrix with potentially non zero entries on the diagonal and off-diagonal \cite{cressie_wls} and can be expected to lead to very precise results. 
However, the computational effort regarding the memory capacities increases at minimum quadratically with the sample size $N$ since the number of pairwise distances that go into the estimation of $\mathbf{\Gamma}$ increases proportionally to $N^2$ making calculations for large N infeasible. %\textcolor{red}{\footnote{elaborated in master thesis}}  

Therefore, the WLS method will be used as next best and applicable solution for the semi-variogram parameter estimation. In case of the WLS estimation method, $\mathbf{\Gamma}$ is a diagonal matrix with varying entries and thus assumes varying variances for each lag's estimate. The covariances between two lags is preset to be zero.
We choose weights depending on the number of data pairs that are considered for the estimation of the empirical semi-variogram according to Matheron's estimator and the middle lag distance assigned to each lag interval. The covariance matrix estimate has entries $\mathbf{\Gamma}_{kk} = h^2_k/|N(h_k)|$ on the diagonal and $\mathbf{\Gamma}_{kl} = 0$ on the non-diagonal for $k,l = 1,\dots, K$ and $k\not = l$. Thereby, minimizing the difference between empirical semi-variogram and model at lag distances with more information 
%that is larger $|N(h_k)|$ 
and at smaller lag distances 
%quantified in terms of $h_k^2$ 
is prioritized.
 The general formula of the loss function to be minimized then reduces to 
\begin{align}\label{min_term_wls}
	LOSS_{\text{WLS}}(\theta) = 
	\sum_{k=1}^K \frac{|N(h_k)|}{h_k^2}\;\cdot [\hat{\gamma(h_k)}-\gamma_{\text{exp}}(h_k;\theta)]^2 
\end{align}
which is implemented in the \verb*|fit.variogram| function from within the Rpackage gstat \cite{pebesma2004gstat} with the default argument \verb*|fit.method=7| \cite{gstatusersmanual}.

\subsection{Generalized bootstrap with filtering}\label{genbootstrap_algorithm}

%The need of uncertainty measurement for semi-variograms was pointed out by \cite{Pardo2001}, \cite{olea2011generalized} already in the geological context. Both articles provide methods for the assessment of semi-variogram model parameter uncertainty estimates.
%But whereas an approximate analytical approach is taken in \cite{Pardo2001}, the empirical approach \cite{olea2011generalized} is used here as a promising starting point since it requires less statistical assumptions regarding the unknown underlying datagenerating spatial process. 
%\\

%The topic of correlation neighbourhood in the health care/epidemiology subject area motivates the exponential semi-variogram modelling based on large sized spatial datasets containing between 500 and 2000 observed subjects \cite{sauzet2021}.
The generalized bootstrap approach \cite{olea2011generalized} serves as starting point, but requires modification to be applicable in the stated data framework of our interest. Each semi-variogram model is fitted using the WLS instead of the GLS method.
Moreover, substantial changes are made with respect to the purpose of the bootstrap estimates. Whereas the original algorithm's goal is the estimation of the covariance matrix and confidence bands of the empirical semi-variogram for each defined lag interval, the focus here lies on the estimation of  model parameter standard errors of a fitted exponential semi-variogram model.
As a completely new element, a filter within the bootstrapping process tackles the exclusion of improbable bootstrap estimates due to non-convergence of the semi-variogram model fitting algorithm.

Considering these aspects, the following algorithm was developed:

\begin{enumerate}
\setcounter{enumi}{-1}
	\item \textbf{Exponential semi-variogram model:} An empirical semi-variogram is estimated based on the original spatial dataset $\mathbf{z} = (z_1,\dots, z_N)^t$ plus the spatial coordinates $\mathbf{s}$ with $\mathbf{s_i} \in \mathbb{R}^2$ for $i = 1,\dots, N$. Subsequently, an exponential semi-variogram model $\gamma_{\text{exp}}$ is fitted using the WLS method producing the parameter estimate $\mathbf{\hat{\theta}}$ for the true parameter vector $\mathbf{\theta} = (c_0, \sigma^2_0, \phi)^t$.

	\item \textbf{Normal score transformation:} The data vector $\mathbf{z}$ is mapped into a Gaussian space by the empirical normal score transformation function $\varphi$ that maps the data from the attribute space into a Gaussian space.
Consequently, $\mathbf{y} = \varphi(\mathbf{z})$ is a realization vector of a standard normal random variable \cite{GSLIB, nscore}.   
	
	\item \textbf{Exponential semi-variogram model for transformed data:} An empirical semi-variogram is fitted  to $\mathbf{y}$ combined with the original data coordinates. Based on that, an exponential semi-variogram model $\tilde{\gamma}_{\text{exp}}$ is fitted using the WLS method providing the parameter estimate $\mathbf{\tilde{\theta}}$.
%In the following parameter uncertainty estimation process, $\tilde{\gamma}_{\text{exp}}$ is used to estimate the covariances between two realisations according to their distance to each other.
%	\item \textbf{Sample's distance matrix:} The distance matrix $\mathbf{D}$ of the sample coordinates is set up as
%\begin{equation}\label{distmatrix}
%\mathbf{D} =
%\begin{pmatrix} 
%  d_{11}     & \cdots  & d_{1N} \\ 
%  \vdots    & \ddots  &  \vdots  \\
%  d_{N1}     & \cdots  & d_{NN}
%\end{pmatrix}
%= 
%\begin{pmatrix} 
%  d(\mathbf{s}_1,\mathbf{s}_1)     & \cdots  &d(\mathbf{s}_1,\mathbf{s}_N) \\ 
%  \vdots    & \ddots  &  \vdots  \\
%  d(\mathbf{s}_N,\mathbf{s}_1)     & \cdots  & d(\mathbf{s}_N,\mathbf{s}_N)
%\end{pmatrix},
%\end{equation}
%where $d(\mathbf{s}_i \; , \mathbf{s}_j)$ denotes the Euclidean distance between two observations located at $\mathbf{s}_i$ and $\mathbf{s}_j$.
	\item \textbf{Covariance estimation:} Making use of the semivarigram model $\tilde{\gamma}_{\text{exp}}$, the covariance between two data points $z_i$ and $z_j$ is calculated based on their Euclidean distance $d_{ij} = d(s_i,s_j)$ as 
\begin{equation}
c_{ij} = \tilde{c_0} + \tilde{\sigma}_0^2 - \tilde{\gamma}_{\text{exp}}(d_{ij}).
\end{equation}
The $N \times N$ covariance matrix with entries $c_{ij}$ is denoted as $\mathbf{C}$.
 
	\item \textbf{Decorrelation of the data:} The decomposition of the covariance matrix $\mathbf{C}$ into the product of the lower triangular matrix $\mathbf{L}$ and its transposed is obtained by the Cholsky decomposition algorithm as
	\begin{equation}
	\mathbf{C} = \mathbf{L}\mathbf{L}^t.
	\end{equation}	
	This is possible because $\mathbf{C}$ is square symmetric and positive-definite \cite{Solow1985}. The vector 
	\begin{equation}
	\mathbf{x} = \mathbf{L}^{-1}\mathbf{y}
	\end{equation}\label{eq_decorrelation}
	is a set of independent and identically distributed, hence uncorrelated values.
	\item \textbf{Classical bootstrap:} Sampling $N$ times with replacement from $\mathbf{x}$ generates a bootstrap sample $\mathbf{x^*}$.
	\item \textbf{Recorrelation:} The bootstrap sample $\mathbf{x^*}$ reinherits the correlation structure by applying the inverse operation of step 4, that is 
	\begin{equation}
	\mathbf{y^*} = \mathbf{Lx^*}.
	\end{equation}		
	\item \textbf{Normal score back transformation:} The back transformation of $\mathbf{y^*}$ to the attribute space through the inverse normal score function in step 1 is done by
	\begin{equation}
	\mathbf{z^*} = \varphi^{-1}(\mathbf{y^*}).
	\end{equation}
	\item \textbf{Analysis of the bootstrap sample:} An exponential semi-variogram model is estimated based on $\mathbf{z^*}$ combined with the original coordinates providing an estimate $\mathbf{\theta^*}$.
\end{enumerate}	

From here on, the algorithm depends on the choice of filter method.	
If the check-based filter is used, steps 9c, 10c and 11 are executed. 
	
\begin{enumerate}
\setcounter{enumi}{8}
\item[(9c)] \textbf{Check-based filtering:} A test is applied to the bootstrap estimate $\theta^*$ to indicate whether the exponential semi-variogram fitting algorithm did converge.
If the algorithm is assumed to have failed, the bootstrap estimate is discarded.
	
	\item[(10c)] \textbf{Repetition:} The steps 5 to 9 are repeated until a set of $B$ bootstrap estimates ${\{ \theta^*_b \}}_{b = 1,\dots, B}$ has aggregated.
\end{enumerate}

If the quantile-based filter is used, steps 9q, 10q and 11 are conducted.

\begin{enumerate}
	\item[(9q)] \textbf{Repetition:} Given the quantile threshold $\alpha \in (0,1)$, the steps 5 to 8 are repeated $\tilde{B} = \frac{1}{\alpha}\cdot B$ times providing a set of $\theta^*_b$ for $b = 1,\dots,\tilde{B}$. 
	\item[(10q)] \textbf{Quantile-based filtering:} The first $B$ values of the increasingly ordered set of bootstrap estimates $\theta^*_b$ is saved. This equals the set of  bootstrap estimates between the minimum value and the $\alpha$-quantile.
	
		\item[(11)] \textbf{Parameter standard error estimation:} Based on the filtered set of bootstrap estimates $\mathbf{\theta}^*_b$ for $b = 1, \dots B$ the parameter standard error estimates are defined as
\begin{align}
\widehat{\eta_{\theta_j}}
= sd(\theta^*_j) 
= \sqrt{ \frac{1}{B-1} \sum_{b=1}^{B}  
\Big\{ \theta_{b j}^* - \overline{\theta_j^*} \Big\}^2 },
\end{align}
for $j = 1,...,3$ referring to the three parameters $c_0, \sigma^2_0, \phi$.
\end{enumerate}

\subsubsection{Check-based filtering}
The check-based filter (steps 9c, 10c) is applied in each repetition. The bootstrap estimate $\theta^*$ is used to compute the corresponding overall variance according to the model as $c_0^* + \sigma_{0}^{2*}$. If
\begin{align}
c_{0}^* + \sigma_{0}^{2*} > \tau Var(\mathbf{z}) \;\; \text{or} \;\; \phi^* < 0,
\end{align}
i.e. if the variance indicated by the model exceeds the sample variance times some threshold factor $\tau$ or the shape parameter lies outside its constraint, the bootstrap estimate $\theta^*$ is discarded. Experience has shown that the WLS algorithm sometimes leads to shape parameter estimates outside its constraint. Otherwise, it is retained. 
%Moreover, experience has shown that the WLS algorithm can sometimes lead to an estimated shape parameter under zero resulting in a no well-defined semi-variogram model by definition. Therefore, if $\phi^* < 0$, the bootstrap estimate is discarded. Otherwise, it is retained.
The size of the bootstrap sample $B$ is specified beforehand and the procedure of estimating and checking one bootstrap estimate is repeated until $B$ bootstrap estimates are accepted.

\subsubsection{Quantile-based filtering}
For each bootstrap estimation, we check whether the parameter constraint $\phi^*> 0$ is complied during the WLS estimation.
After the computation of all bootstrap estimates the quantile-based filter (steps 9q, 10q) is applied.
Given a threshold percentile $\alpha$ all bootstrap estimates $\theta^*_b$ beyond the $\alpha\cdot100\%$-quantile $q_\alpha(\theta^*_j)$ for $j=1,2,3$ are discarded while all bootstrap estimates up to the $\alpha\cdot100\%$-quantile are saved.

To make sure the set of bootstrap estimates used for the computation of the parameter standard error contains $B$ elements, the number of bootstrap samples to be generated in the first place is set to $\tilde{B} = B \cdot \frac{1}{\alpha}$.

\section{Simulation study}\label{simstudysection}

A simulation study is conducted in order to test and evaluate the performance of the presented filter methods in various sampling frameworks and provide recommendations for meta- and tuning parameter specification.
The implementation is constructed using the statistical software environment R \cite{R}. The R package gstat \cite{pebesma2004gstat} is used to simulate spatial data. The packages gstat \cite{pebesma2004gstat}, SpatialTools \cite{SpatialTools} and additional individual functions \cite{nscore} are used for the implementation of the generalized bootstrap algorithm with each filter method. The algorithms and simulation code are stored within the Github repositories EgoCor \cite{EgoCorGithub} and EgoCorSim \cite{EgoCorSimGithub}.
% eventuell auf Supplementary material verweisen

\subsection{Data generating process}
Data is generated based on a zero-mean Gaussian random process with an exponential semi-variogram model of the form
\begin{align}
\gamma(h) 
= 60 + 40\cdot \Big\{1 - \exp\Big(- \frac{h}{200}\Big)\Big\} \;\; & \text{for}\;\; h > 0.
\end{align}
The population is located on a two-dimensional square grid $D = \{0,\dots,10000\}^2$. One complete realization of locations $(\mathbf{x,y})$ is generated by drawing 250000 data points using the \verb !runif! command. The attribute values $z$ at each realized location are computed with the following commands.
\begin{verbatim}
  model = gstat::vgm(nugget = 60, psill = 40, range = 200, model = "Exp")
  true.model = gstat::gstat(formula = z~1, locations =~x+y, dummy = T,        
                            beta = 0, model = model, nmax = 100)
  cr = stats::predict(true.model, newdata = xy, nsim = 1)
\end{verbatim}

\subsection{Sampling schemes}\label{pars_sampling}
\begin{figure}
	{\centering
	\includegraphics[width = \textwidth]{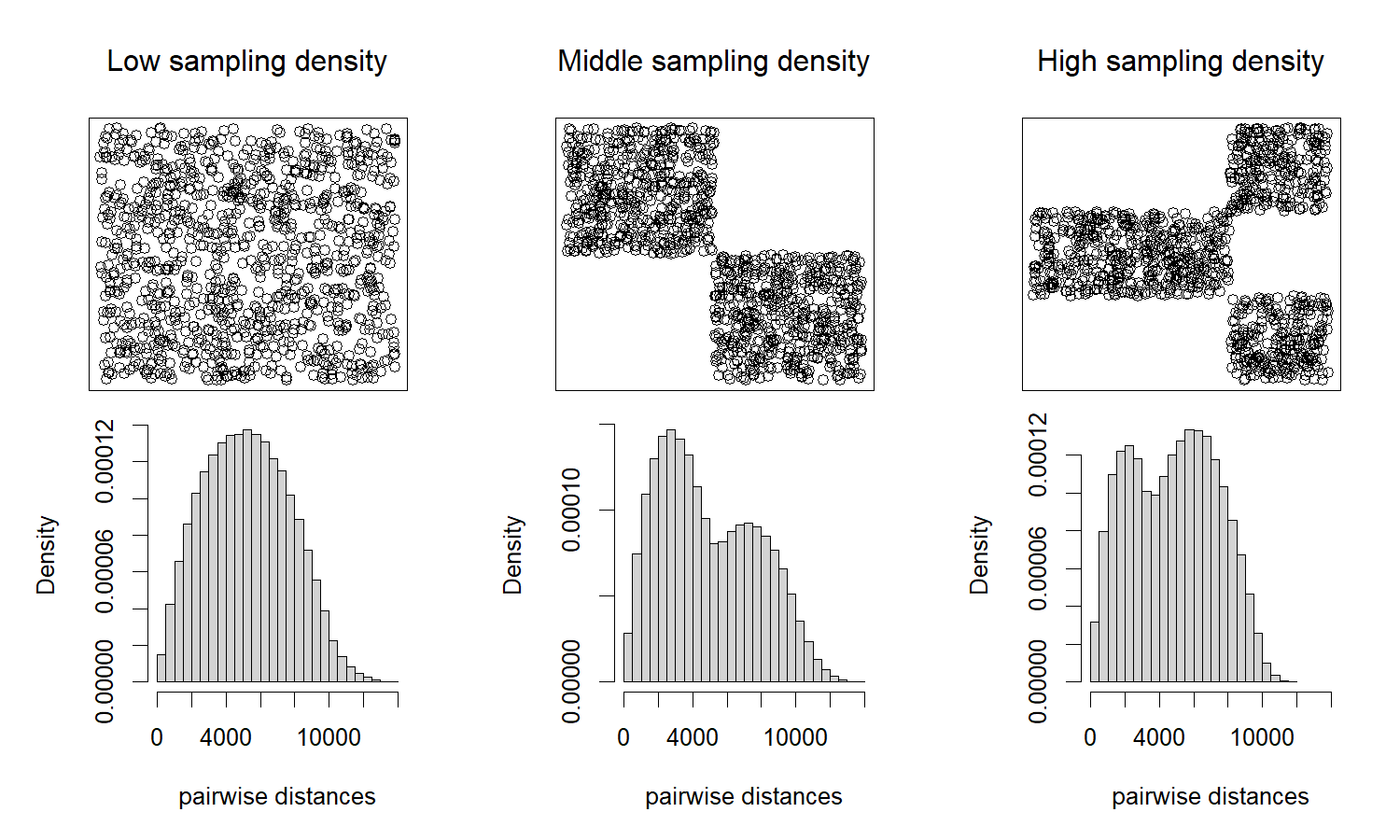}}
	\caption{Example for sampling densities and resulting distribution of pairwise distances.
	Three subsets are sampled from the simulated population with $N = 1000$ and either a low, middle or high sampling density. The upper row shows the sample locations on the map $D$ according to the sampling density schemes. The empirical distribution of the pairwise distances is provided below.}
	\label{fig:sampling_example}
\end{figure}

Samples are produced by sampling from the simulated population without replacement using the \verb!sample! function. Sample scenarios vary with respect to the sample size $N$ ($500, 1000, 2000$) and the sampling density  (low, middle, high). In case of a low density, the sampled locations are uniformly distributed over the whole grid surface. Consequently, many point pairs with similarly valued pairwise distances occur in the middle of the pairwise distances range. In comparison, a high sampling density provides more small valued pairwise distances (Figure \ref{fig:sampling_example}).

\subsection{Metaparameters for the empirical semi-variogram}\label{pars_sv}
%The empirical semi-variogram is obtained by the \verb!variog! command and the outcome is then inserted into the \verb!variofit!. For the estimation process two metaparamters have to be specified.

The maximal distance specifies the maximum value up to which pairwise distances are considered within the semi-variogram estimation process.
Three values $1.1\cdot h^*, 1.5\cdot h^*$ and $2\cdot h^*$ are used. The true practical range according to the given semi-variogram model is $h^* = 415.89$.
%calculated as $h^* = ln(\frac{\sigma_0^2}{0.05(c_0 + \sigma^2_0)})\cdot \phi$

The number of lag intervals is chosen to be $10$. Accordingly, the range of pairwise distances from 0 to the maximal distance is divided into 10 equidistant lag intervals. For each interval, a point estimate is calculated based on all point pairs within the lag bin for the empirical semi-variogram.

\subsection{Tuning parameters for the filter methods}\label{pars_filter}
For the check-based filter approach, the factor $\tau$ scaling the threshold of the test has to be specified. Within the frame of the simulation study, the values $1.1,1.2,1.5,2.0,2.5,3.0$ are applied.

For the quantile-based filter method, the percentage of the quantile threshold $\alpha$ has to be chosen. Values ranging from $0.75$ to $1.0$ with a step length of $0.05$ are put to the test. The value 1.0 equals the case with no filtering and serves as benchmark.

\subsection{Conduction of the simulation study and elimination of ill-fitting models}\label{sec:improb_runs}

Each scenario is simulated $n_{sim} = 3000$ times.
Very large or non-positive estimates for at least one of the semi-variogram parameters are an indicator that the algorithm did not converge in the estimation process. 
%While a first impression of the estimated semi-variogram model is obtainable by eye in practice, the framework of the simulation study requires an automatic check. For this purpose, every estimated model fulfilling the rule
Only estimated models fulfilling
\begin{align}
0 < \hat{c_0}  < 1000 \;\; \text{and} \;\;
0 < \hat{\sigma_0^2}  < 1000 \;\; \text{and} \;\;
0 < \hat{\phi}  < 1000
\end{align}
are assumed to have converged and will be included into the evaluation. This leaves $\tilde{n}_{sim} < n_{sim}$ simulation runs to be analyzed. The proportion  $\tilde{n}_{sim}/n_{sim}$ will be referred to as the convergence rate in section \ref{sec:results}.

\subsection{Performance measures}
The performance statistics are presented as averages grouped by the sample size, the sampling density, the maximal distance factor and the tuning parameter specification for each filter method.
The empirical standard deviation over the parameter estimates of all simulation runs $\eta_{\theta_j} = sd(\hat{\theta_j})$ for $j=1,2,3$ serves as approximation to the true standard error and benchmark value for our performance evaluation.
Its variability is obtained by the Monte Carlo standard error estimator $se_{MC}(\eta_{\theta_j}) = sd(\hat{\theta_j})/\sqrt{2(\tilde{n}_{sim} -1)}$. 
The performance will be evaluated in terms of the estimates' standard deviation, bias and mean squared error.

The grouped average estimate $\overline{\widehat{\eta_{\theta_j}}}$ according to each grouping parameter is provided along with its grouped average of the corresponding standard deviations $sd(\widehat{\eta_{\theta_j}})$.
For instance, the grouped average estimate given the grouping factor $N = 500$ is obtained by averaging over all estimates for sampling scenarios with $N=500$ and arbitrary specification for the sampling density, maximal distance factor and tuning parameter. The standard deviations of the same scenarios are averaged to obtain the grouped average standard deviation.

The bias is defined as 
\begin{align}
bias(\widehat{\eta_{\theta_j}}) =
\overline{\widehat{\eta_{\theta_j}}} - \eta_{\theta_j}.
\end{align}
The overall performance is measured in terms of the mean squared error (MSE)
\begin{align}
MSE(\widehat{\eta_{\theta_j}}) = sd(\widehat{\eta_{\theta_j}})^2 + bias(\widehat{\eta_{\theta_j}})^2
\end{align}
taking into account the performance in regards to both, the location of the estimated standard error and the variability described by the squared empirical standard deviation of the estimated standard error estimates.

\section{Results}\label{sec:results}
In the following, the simulation study results are described. Compact result tables can be found in the appendix. %Observed statistics are provided in percentage with one decimal place or as decimal number rounded to the second decimal place. Large numbers ($> 10^3$) are presented in exponential notation for better readability.

\subsection{Convergence rate for the estimation of the parameters }%{\footnote{Of what?}

The convergence rate $\tilde{n}_{sim}/n_{sim}$ is on grouped average lowest for a sample size of 500 with $51\%$. It increases with increasing sample size leading to $76\%$ for $N = 1000$ and $92\%$ for $N = 2000$.
An increase in the sampling density leads to a slighter increase in the convergence rate ranging from $67\%$ for a low sampling density to $78\%$ in case of middle and $95.9\%$ in case of a high sampling density.
Similarly, an increase in the maximal distance factor provides a slighter increase in the convergence rate from $61\%$ for a maximal distance factor of $1.1$ over $75\%$ for a maximal distance factor of $1.5$ to $84\%$ for a maximal distance factor of $2$ (Table \ref{table_prob_runs}).

\subsection{Performance of generalized bootstrap approach with no filter}

%TABLE with no filter results
\begin{table}
\centering
\parbox{\textwidth}{\caption{Generalized bootstrap method with no filter - Performance measure results for the nugget effect $ c_0 $, partial sill $\sigma_0^2$ and shape parameter $\phi$. Result statistics are rounded to the 2nd decimal point and more for numbers with more than 6 decimals.}} 
\label{perftable_no_filter}
\begingroup\footnotesize
\begin{tabular}{|L{2cm}R{3cm}R{3cm}R{1.5cm}R{2.5cm}|}
  \hline
semi-variogram parameter & $\eta_{\theta_i}$ ($se_{MC}(\eta_{\theta_i})$) & $\overline{\widehat{\eta_{\theta_i}}}$ ($sd(\widehat{\eta_{\theta_i}})$) & $bias(\widehat{\eta_{\theta_i}})$ & $MSE(\widehat{\eta_{\theta_i}})$ \\ 
  \hline  
  $c_0$ & 14.95 (0.0433) & 6.64 $\;\;\;\;\;\;\;\;\;$(4.21) & -8.31 & 86.75 \\ 
  $\sigma_0^2$ & 14.61 (0.0424) & 1606.99 $\;\;$(55093.93) & 1592.38 & $3.04\cdot 10^9 \;$ \\
  $\phi$ & 166.43 (0.4827) & 6035.77 (184498.18) & 5869.34 &  $3.41\cdot 10^{10}$ \\ 
   \hline
\end{tabular}
\endgroup
\end{table}

The standard error estimates calculated with the generalized bootstrap approach with no filter for nugget effect, partial sill and shape parameter averaged with respect to the implemented sampling schemes and metaparameters differ in the observed performance measurements (see Table \ref{perftable_no_filter}).
For nugget effect standard errors, small standard deviation and slightly negative bias is obtained. The corresponding MSE of 86.75 is relatively small.
For the partial sill and shape parameter we observe a high variance and high positive bias. This results in a large MSE of more than $3\cdot10^9$ for both parameters.

\subsection{Performance of the check filter method}
In the following, the effects of the sample size, sampling density, maximal distance and choice of threshold factor on the performance of the standard error estimation with check filtering are presented based on the grouped average outcomes.
Tables containing the detailed results of the simulation study results can be found in the appendix (Tables \ref{perftable_check_nugget}, \ref{perftable_check_partial sill}, \ref{perftable_check_shape}).
\\

\textbf{\textit{Effect of sample size:}} A higher sample size improves all parameters' standard error estimates. This is indicated by the decreasing MSE as overall performance measure observed for increasing sample sizes. % Exceptions to this tendency are observed in the bias of $\eta_{\sigma_0^2}$ and $\eta_{\phi}$ which is smallest in case of a sample size of 500...
\\

\textbf{\textit{Effect of sampling density:}}
For $\widehat{\eta_{c_0}}$ and $\widehat{\eta_{\phi}}$ all performance measurements indicate very slight improvement with increasing sampling density. For $\widehat{\eta_{\sigma_0^2}}$ the estimation is relatively worst in case of the middle sampling density, improves little in case of the high sampling density and best when using the small sampling density.
It is not clear whether the small differences between low, middle and high sampling density results indicate an effect at all.
\\

\textbf{\textit{Effect of maximal distance:}}
Specifying a higher maximal distance for the model fitting improves the standard error estimation slightly for all semi-variogram parameters. This effect is observed in all performance measurements.
\\

\textbf{\textit{Effect of tuning parameter:}}
Although, the check filter has the expected effect of reducing the variance (or standard deviation $sd(\widehat{\eta_{\theta_i}})$ respectively) and the bias of the standard error estimation of all parameters, the effect on the performance and thereby the threshold factor value optimizing the estimation performance is different for each. 

The standard error for the nugget effect is estimated best when using the generalized bootstrap approach with no filter. When applying the check filter, a smaller $\tau$ (equivalent to more severe filtering) leads to more underestimated standard error and thereby to a larger MSE. However, the observed effect is very small.

Adding a check filter to the partial sill standard error estimation improves the performance immensely compared to the generalized bootstrap estimation with no filtering. Among the implemented threshold  factors, we observe that decreasing $\tau$ leads to better estimation results according to all three performance measures.

Similar to the partial sill, a check filter applied in the shape parameter standard error estimation leads to strongly improved estimator performance compared to the generalized bootstrap approach with no filtering.
Among the implemented threshold factors, slight differences in the performance measurements are observed.
While most severe filtering with $\tau=1.1$ minimizes the variance in the estimations, simultaneously, the bias is closest to zero for $\tau = 3.0$. More severe filtering leads to more negative bias. According to the MSE, $\tau=1.5$ is a suitable compromise.
\\

\subsection{Performance of the quantile filter method}

The effects of the sample size, sampling density, maximal distance and choice of quantile threshold value on the performance of the standard error estimation with quantile filtering are presented based on the grouped average outcomes.
Tables containing the detailed simulation study results can be found in the appendix (Tables \ref{perftable_quantile_nugget}, \ref{perftable_quantile_partial sill}, \ref{perftable_quantile_shape}).
\\

\textbf{\textit{Effect of sample size:}}
Increasing the sample size leads to an improvement of the standard error estimation for all three parameters according to all performance measures.
\\

\textbf{\textit{Effect of sampling density:}}
For $\eta_{c_0}$, a slight improvement in the estimation performance is observed when the sampling density is higher. 
For $\widehat{\eta_{\sigma_0^2}}$ and $\widehat{\eta_{\phi}}$, a middle to high sampling density leads to better estimator performance compared to a low sampling density with slight preference for the middle density in all performance measures.
\\

\textbf{\textit{Effect of maximal distance:}}
The effect of the maximal distance on the estimation performance shows the same tendencies as the effect of the maximal distance.
For $\widehat{\eta_{c_0}}$, a slight improvement in the estimation performance is observed when the maximal distance is increased. 
For $\widehat{\eta_{\sigma_0^2}}$ and $\widehat{\eta_{\phi}}$, a maximal distance factor of 1.5 or 2.0 leads to better estimator performance compared to a maximal distance factor of 1.1 with slight preference for the 1.5 factor in regards to all performance measures.
\\

\textbf{\textit{Effect of tuning parameter:}}
As expected, more severe filtering by smaller quantile threshold values $\alpha$ decreases variance and bias of the standard error estimator for each parameter. But this has different implications on the performance for each parameter.

For the nugget effect, standard error estimation using the generalized bootstrap without filter leads to slightly less negative bias than applying a filter. Comparison of the MSE as overall performance measure implicates the same: using no filter leads to the best standard error estimation for the nugget effect.

For the partial sill, using a filter improves the standard error estimation by far compared to the approach without filtering. Although the differences in performance are much smaller when comparing the results for the different implemented quantile thresholds, a slight trend toward smaller $\alpha$ values and thereby to more severe filtering within the bootstrap to improve performance according to all performance measures is observed.

For the shape parameter, the addition of the filter in the generalized bootstrap leads to immense improvement within the standard error estimation compared to the no-filter version. Although even the least severe filter with $\alpha=0.95$ leads too a small negative bias, the MSE indicates that more severe sampling improves the standard error estimation in regards to bias and variance considered together.

\begin{figure}
	{\centering
	\includegraphics[width = \textwidth]{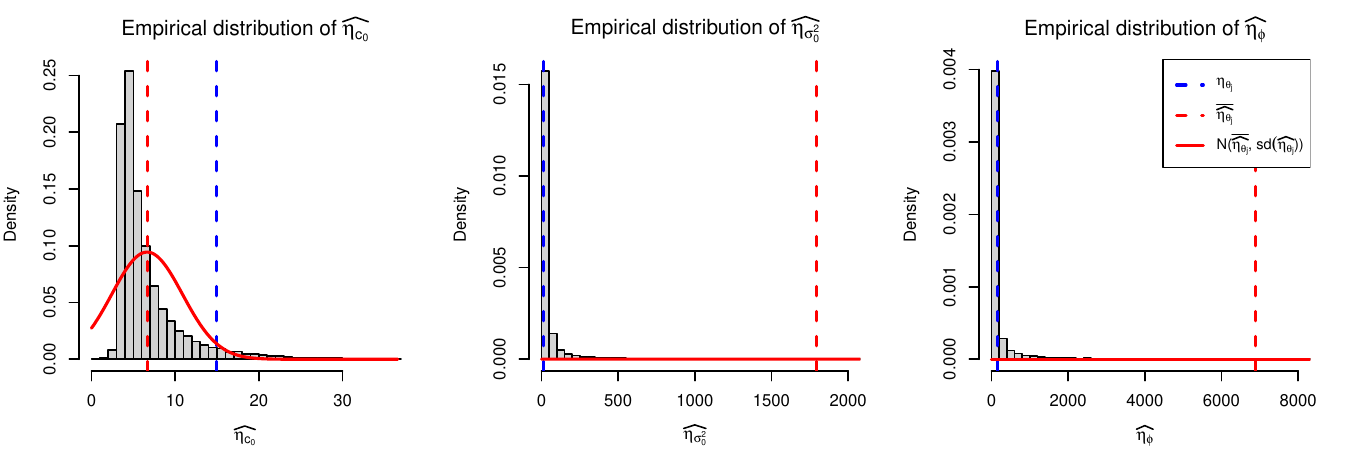}
	\includegraphics[width = \textwidth]{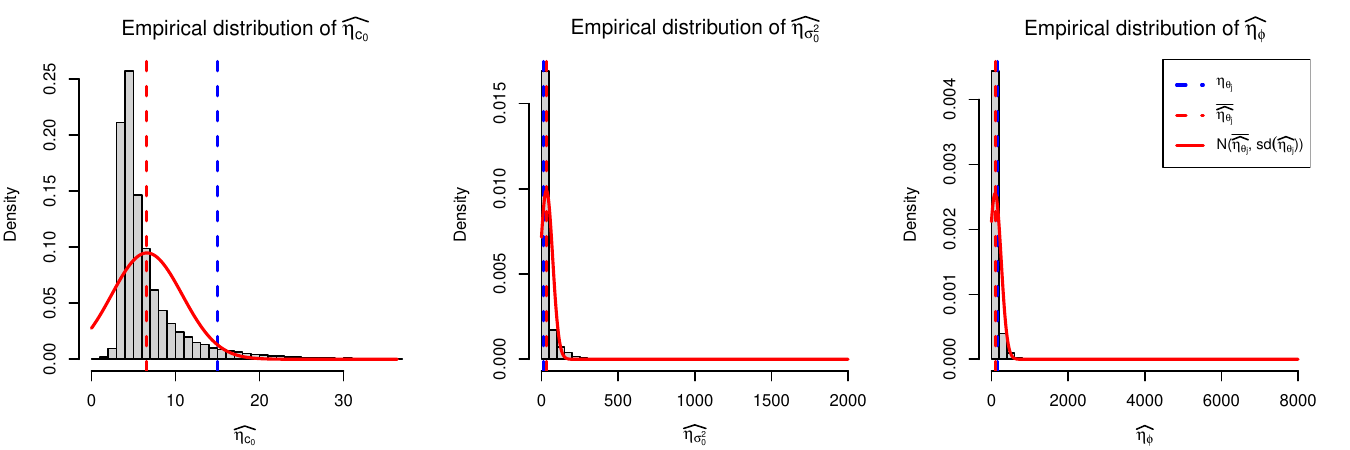}}
	\caption{Empirical distribution of standard error estimates obtained with generalized bootstrap without filtering (row 1) and with check filtering with threshold value $3.0$ (row 2).
	Each histogram is based on all simulation results (all framework conditions). The standard deviation calculated from all simulation estimates $\eta_{\theta_i}$ is marked in blue. The average standard error estimate $\overline{\widehat{\eta_{\theta_i}}}$ with normal density based on the average standard error estimate and standard deviation of the standard error estimates are provided in red.
}
	\label{fig:sim_emp_dists_ses}
\end{figure}

\section{Discussion}

The generalized bootstrap approach of Olea and Pardo-Igúzquiza \citep{olea2011generalized} provides standard error estimates which our simulation study shows to be hugely biased upwards and rather instable as far as  the standard error estimator of the partial sill and shape parameter are concerned (see Figure \ref{fig:sim_emp_dists_ses}). An explanation can be found in the semi-variogram estimation in praxis: 
\begin{enumerate}
\item The data analyst fits multiple semi-variogram models trying multiple metaparameter values, for instance for the number of bins and maximal distance.
\item The data analyst evaluates by eye how well the semi-variogram model fits the corresponding empirical semi-variogram. The best looking model is chosen for further statistical inference. The other models are discarded. 
\end{enumerate}

This procedure contains implicit filter mechanisms. Ill-fitting models and corresponding parameter estimates are thus discarded and we need to introduce an automated filter mechanism within each bootstrap repetition. Therefore, we developed explicit check- and quantile filter rules and tried a range of tuning alternatives within the simulation study. We selected the best alternative based on the simulation results. \\

The issue of ill-fitting parametric semi-variograms concern also the simulation study itself. For that reason, we implemented the heuristic filter rule stated in sec. \ref{sec:improb_runs} to identify and exclude generated data that would supposedly be discarded by a practitioner. However this method is imperfect and makes the evaluation of the "true" standard error of the semi-variogram parameters difficult.

\subsection{The optimal filter}
Compared to the generalized bootstrap approach without filter, all implemented filter extensions reduce the bias upwards and the instability in the standard error estimates. 
While applying any filter has a negligible negative effect on the performance of the nugget effect's standard error estimator relative to the true standard error, estimates of standard errors for the partial sill and shape parameters are immensely improved by the procedure.
\\
The check and quantile filter extension are similar in that they exclude bootstrap parameter estimates that are too high. An advantage of the check-based filter approach is the automated comparison of the spatial data variance with the variance estimated according to the semi-variogram model $c_0+\sigma_0^2$. 
Due to this dependence, the check filter is mainly regulated by the empirical variance induced by the data. The threshold factor provides an option for fine tuning and can be calibrated with the simulation results. The lack of relation between the quantile-based filter and the data context makes tuning of the corresponding tuning parameter $\alpha$ less intuitive.

\subsection{Effects of simulation parameters on the standard error estimation}
The effects of sample size, sampling density and maximal distance specification on the performance of the standard error estimation observed in the simulation study mirror the effect on parameter estimation of the semi-variogram itself as  seen in Sauzet et al. \cite{sauzet2021}. Because of the empirical nature of the bootstrap approach, adjusting the framework parameters to improve semi-variogram modelling automatically improves the performance of the standard error estimation.

\subsection{Recommendations for action} 
We recommend the generalized bootstrap method with check-based filtering and tuning parameter specification $\tau = 3.0$. If controllable, sample size and density are preferably large in order to provide enough point pairs within the lag bins up to the practical range.

\section{Case study}

\begin{figure}
\centering
%\subfloat[An example of an individual figure sub-caption.]{%
\resizebox*{0.5\linewidth}{!}{\includegraphics{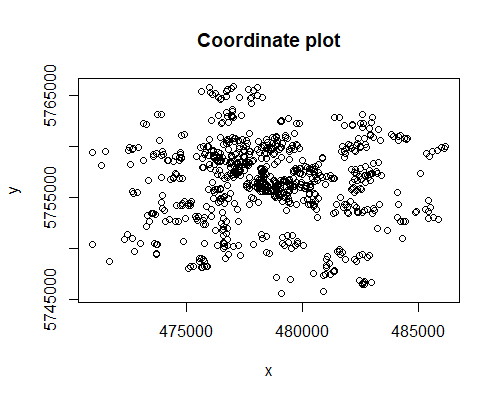}}%}
%\hspace{5pt}
%\subfloat[A slightly shorter sub-caption.]{%
\resizebox*{0.5\linewidth}{!}{\includegraphics{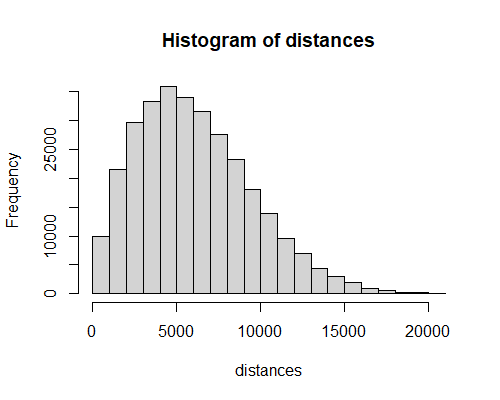}}%}
\caption{Plot of geo-masked coordinates (left) and a histogram of all pairwise Euclidean distances resulting from the coordinates (right).} \label{fig:case_study_coordinate_plot}
\end{figure}

The generalized bootstrap with check-based filtering is demonstrated on data obtained from the BaBi birth cohort study \cite{babi_protocol}. Within the study period from 2013 to 2016 information were gathered about 965 newborn babies and their families including health status attributes and household locations.

Following the same steps as in the work of Sauzet et al. we use geo-masked coordinates of the household locations to model the spatial semi-variance of birthweights  \cite{sauzet2021}. Excluding subjects with incomplete information leaves 783 data points for the modelling procedure. The geo-masked locations of the households and a histogram of the pairwise distances are shown in Figure \ref{fig:case_study_coordinate_plot}.

Besides the choice of the exponential semi-variogram model form, the maximal distance and number of lag intervals have to be chosen. We fit models with varying metaparameter specifications. The fit is evaluated by visual inspection and comparison of the estimated variance according to the semi-variogram model $\hat{\sigma}^2_{model} = \hat{c_0} + \hat{\sigma}_0^2$ with the sample variance $\hat{\sigma}_{sample}^2$ by calculating the relative bias as the ratio
\begin{align}
	RB = \frac{\hat{\sigma}^2_{model}}{\hat{\sigma}_{sample}^2} .
\end{align}

First, we try out a range of maximal distances $(400,500,600,700,800,900,1000)$ and one number of lags $(13)$.
By looking at visualizations of the models, we tend to a maximal distance of $800$. The estimated variance is higher by a factor of $2.09\%$ than the sample variance according to the relative bias. 

Given the maximal distance of $800$ we try out multiple lag bin numbers $(10,11,12,13,14,15)$.
Visually, the models with 10 lags (Model 1) and 13 lags (Model 2) seem to have similarly good fit and provide the two smallest relative biases by a factor of $1.16\%$ and $2.09\%$ out of all alternative models.

We estimate the parameter standard errors using the generalized bootstrap with check-based filtering and tuning parameter $\tau = 3.0$.
While the parameter estimates for both models are similar, Model 2 provides smaller standard error estimates and is selected as final model (Table \ref{tab_birthweight_estimates}). 

\begin{figure}
\centering
\includegraphics[width = \textwidth]{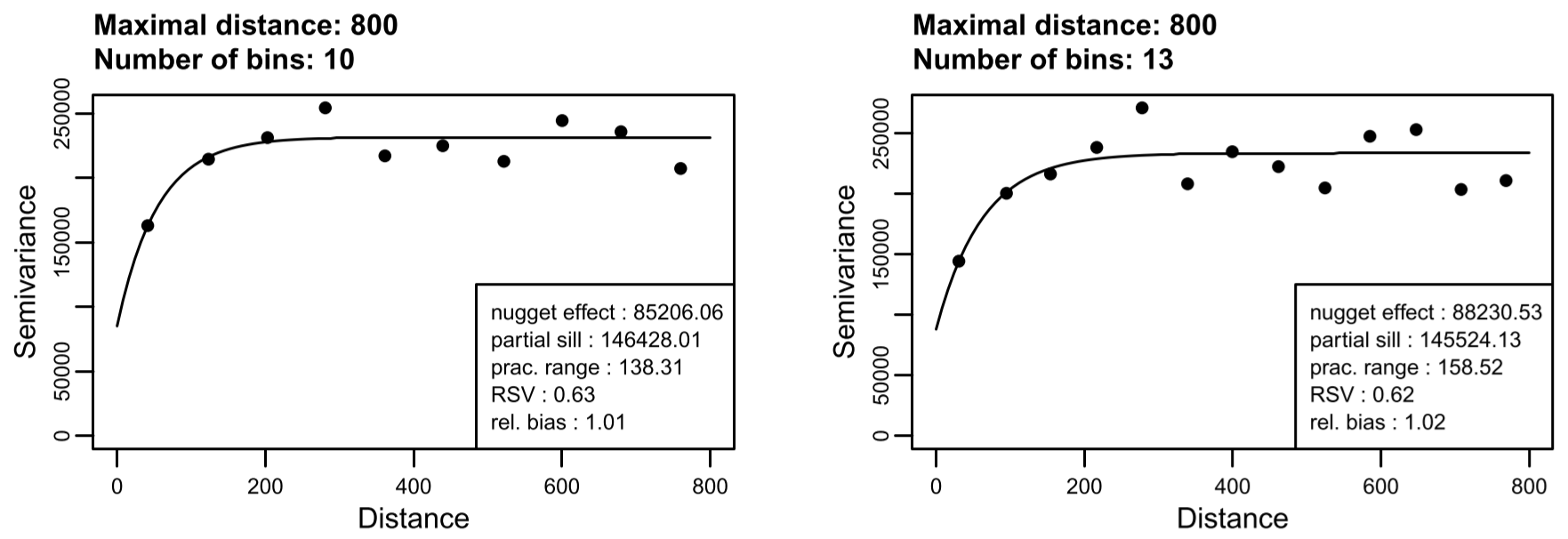}
\caption{Semi-variogram models fitted to birthweight data with a prespecified maximal distance of 800 and 10 lag intervals (left), as well as, 13 lag intervals (right).} \label{fig:bithweight_sv_mod}
\end{figure}

\begin{table}
	\caption{Semi-variogram model parameter and standard error estimates for birthweight data. Model 1 refers to the exponential semi-variogram model with a prespecified maximal distance of 800 and 10 lag intervals. Model 2 refers tothe exponential semi-variogram model with a prespecified maximal distance of 800 and 13 lag intervals.} 
	{\begingroup\footnotesize
		\begin{tabular}{|lR{2cm}R{2.5cm}R{2cm}R{2.5cm}|}
			\hline
			&&& & \\
			& \multicolumn{2}{c}{\textbf{Model 1}} & \multicolumn{2}{c|}{\textbf{Model 2}} \\
			& \textbf{Parameter estimate} & \textbf{Standard error estimate} &  \textbf{Parameter estimate} & \textbf{Standard error estimate}\\
			%	\hline
			&&&& \\
			$\mathbf{c_0}$       & 85206.06 & 141638.53 & 88230.53 & 145214.64 \\
			$\mathbf{\sigma_0^2}$ & 146428.01 & 366958.70 &145524.13 & 338533.10 \\
			$\mathbf{\phi}$    &  54.51 & 419.08 &  62.86 & 337.79 \\
			&&&& \\
			
			\hline
		\end{tabular}\endgroup}
	%\tabnote{\textsuperscript{a}This footnote shows how to include footnotes to a table if required.}
	\label{tab_birthweight_estimates}
\end{table}

Estimates of semi-variogram parameters have very skewed distributions and the standard deviation cannot be interpreted in the usual sense. Here, the large values are to be understood as how unstable the model is and indicate that the evidence of spatial correlation in the data is rather poor. This follows from the limited number of pairs of observations at small-distances.

\section{Conclusion and limits}

\subsection{Conclusion}
Standard errors for the exponential semi-variogram model parameters provide a basis for inference and evaluation of the fitted model that exceeds subjective evaluation by eye.
Existing approaches do not work reliably for typical social epidemiological spatial analyses.
We developed new approaches to estimate the standard errors of the parameters of an exponential semi-variogram model: The generalized bootstrap method with either check- or quantile-based filtering. Both approaches were presented and evaluated using a simulation study considering multiple framework scenarios varying with regards to sample size, sampling density and maximal distance specifications, as well as multiple tuning alternatives for each filter option.
According to the simulation study results, all implemented filter extensions lead to a massive improvement of the standard error estimator compared to the generalized bootstrap without filtering. Another pro argument for the filter extensions is the fact that the implemented filter serves as heuristic rule to approximate the implicit filter mechanism conducted during semi-variogram model fitting in practice. The check filter has a stronger relation to the data variance and is therefore believed to be more intuitive and easier to tune than the quantile filter.
Based on the simulation study results and discussed thoughts, we recommend the generalized bootstrap method with check-based filtering and threshold factor $\tau = 3.0$ for application.
%The semi-variogram modelling workflow supposedly contains an implicite filter mechanism with respect to the hypothetical set of fitted semi-variogram models and justifys the implementation of a filter into the bootstrap method.

\subsection{Limits}
So far, we focused on the exponential semi-variogram with nugget effect since it is of great interest in the study of neighbourhood correlation. However, the generalized bootstrap method with filtering is conceptualized for every kind of parametric semi-variogram model, such as Gaussian, Whittle or Spherical. A simulation study for these parametric model forms has to be conducted, yet.

%--------------------------------------template sections

\backmatter

\bmhead{Acknowledgments}

We thank all contributors and authors of the BaBi cohort study \cite{babi_protocol} for the permission to use the dataset.

\bmhead{Disclosure statement(s)}
The authors report there are no competing interests to declare.

\bmhead{Data availability statement}
Data from the BaBi study was used for demonstration purposes. Because of ethical restrictions, the data are available only upon request.

\bibliography{articlerefs}% common bib file

%%===================================================%

\begin{appendices}

\section{Simulation study results}

\begin{table}[h!]
\caption{Percentage of averaged probable runs grouped by sample size, sample density and maximal distance factor. The number of runs per grouping parameter combination (e.g. sample size of 500, low sample density, maximal distance factor of 1.1) is $n_{sim} = 3000$. The number of all executed runs per group is then $27000$ and equals 100\%. Parameter estimates are obtained by the WLS approach (sec. \ref{modelfitting}).} 
{\begingroup\footnotesize
\begin{tabular}{|L{3cm}R{2cm}R{2cm}R{3cm}|}
		\hline
		grouping parameter & $n_{sim}$ & $\tilde{n}_{sim}$ & \% probable WLS runs \\ 
		\hline 
		\multicolumn{4}{|l|}{\textbf{$\varnothing$ by sample size}} \\ 
		500 & 27000 & 13901 & 51\%  \\ 
		1000 & 27000 & 20592 & 76\%  \\ 
		2000 & 27000 & 24962 & 92\%   \\ 
		\hline 
		\multicolumn{4}{|l|}{\textbf{$\varnothing$ by sample density}} \\ 
		low & 27000 & 18058 & 67\%   \\ 
		middle & 27000 & 20438 & 76\%   \\ 
		high & 27000 & 20959 & 78\%   \\ 
		\hline 
		\multicolumn{4}{|l|}{\textbf{$\varnothing$ by maximal distance factor}} \\ 
		1.1 & 27000 & 16488 & 61\%   \\ 
		1.5 & 27000 & 20276 & 75\%   \\ 
		2.0 & 27000 & 22691 & 84\%   \\ 
\hline
	\end{tabular}\endgroup}
%\tabnote{\textsuperscript{a}This footnote shows how to include footnotes to a table if required.}
\label{table_prob_runs}
\end{table}

% latex table generated in R 4.3.0 by xtable 1.8-4 package
% Mon Jun 12 15:47:26 2023
\begin{table}[h!]
\centering
\parbox{\textwidth}{\caption{Check Filter Method - Performance measure results for the nugget effect $ c_0 $ grouped by sample size, sample density, maximal distance factor and tuning parameter $\tau$.}} 
\label{perftable_check_nugget}
\begingroup\footnotesize
\begin{tabular}{|L{2cm}R{3.5cm}R{3cm}R{2cm}R{2cm}|}
  \hline
semi-variogram parameter & $\eta_{\theta_i}$ ($se_{MC}(\eta_{\theta_i})$) & $\overline{\widehat{\eta_{\theta_i}}}$ ($sd(\widehat{\eta_{\theta_i}})$) & $bias(\widehat{\eta_{\theta_i}})$ & $MSE(\widehat{\eta_{\theta_i}})$ \\ 
  \hline 
 \multicolumn{5}{|l|}{\textbf{$\varnothing$ grouped by sample size}} \\ 
500 & 19.62 (0.1177) & 10.11 (5.6337) & -9.5069 & 122.1197 \\ 
  1000 & 15.08 (0.0743) & 6.59 (3.2210) & -8.4942 & 82.5259 \\ 
  2000 & 8.58 (0.0384) & 4.08 (1.0185) & -4.5009 & 21.2949 \\ 
   \hline 
 \multicolumn{5}{|l|}{\textbf{$\varnothing$ grouped by sample density}} \\ 
low & 16.63 (0.0875) & 6.87 (4.6803) & -9.7598 & 117.1596 \\ 
  middle & 14.15 (0.0700) & 6.10 (3.8460) & -8.0458 & 79.5263 \\ 
  high & 14.01 (0.0684) & 6.17 (3.7961) & -7.8395 & 75.8688 \\ 
   \hline 
 \multicolumn{5}{|l|}{\textbf{$\varnothing$ grouped by maximal distance factor}} \\ 
1.1 & 15.58 (0.0858) & 6.22 (4.6159) & -9.3582 & 108.8821 \\ 
  1.5 & 14.71 (0.0731) & 6.24 (4.0820) & -8.4708 & 88.4165 \\ 
  2.0 & 14.36 (0.0674) & 6.57 (3.7317) & -7.7848 & 74.5286 \\ 
   \hline 
 \multicolumn{5}{|l|}{\textbf{$\varnothing$ grouped by tuning parameter $\tau$}} \\ 
1.1 & 14.95 (0.0433) & 5.86 (3.8942) & -9.0828 & 97.6615 \\ 
  1.2 & 14.95 (0.0433) & 6.17 (4.0354) & -8.7775 & 93.3292 \\ 
  1.5 & 14.95 (0.0433) & 6.44 (4.1440) & -8.5123 & 89.6325 \\ 
  2.0 & 14.95 (0.0433) & 6.54 (4.1768) & -8.4112 & 88.1942 \\ 
  2.5 & 14.95 (0.0433) & 6.57 (4.1877) & -8.3764 & 87.7004 \\ 
  3.0 & 14.95 (0.0433) & 6.59 (4.1931) & -8.3590 & 87.4549 \\ 
   \hline
\end{tabular}
\endgroup
\end{table}

% latex table generated in R 4.3.0 by xtable 1.8-4 package
% Mon Jun 12 15:50:02 2023
\begin{table}[h!]
\centering
\parbox{\textwidth}{\caption{Check Filter Method - Performance measure results for the partial sill effect $ \sigma^2_0 $ grouped by sample size, sample density, maximal distance factor and tuning parameter $\tau$.}} 
\label{perftable_check_partial sill}
\begingroup\footnotesize
\begin{tabular}{|L{2cm}R{3.5cm}R{3cm}R{2cm}R{2cm}|}
  \hline
semi-variogram parameter & $\eta_{\theta_i}$ ($se_{MC}(\eta_{\theta_i})$) & $\overline{\widehat{\eta_{\theta_i}}}$ ($sd(\widehat{\eta_{\theta_i}})$) & $bias(\widehat{\eta_{\theta_i}})$ & $MSE(\widehat{\eta_{\theta_i}})$ \\  
  \hline 
 \multicolumn{5}{|l|}{\textbf{$\varnothing$ grouped by sample size}} \\ 
500 & 19.68 (0.1180) & 30.63 (31.8383) & 10.9492 & 1133.5601 \\ 
  1000 & 13.76 (0.0678) & 26.44 (30.0340) & 12.6813 & 1062.8580 \\ 
  2000 & 8.63 (0.0386) & 23.56 (24.6203) & 14.9233 & 828.8651 \\ 
   \hline 
 \multicolumn{5}{|l|}{\textbf{$\varnothing$ grouped by sample density}} \\ 
low & 16.24 (0.0855) & 23.03 (26.6813) & 6.7810 & 757.8730 \\ 
  middle & 13.84 (0.0685) & 28.43 (29.7273) & 14.5837 & 1096.3960 \\ 
  high & 13.7 (0.0669) & 26.79 (28.5118) & 13.0932 & 984.3554 \\ 
   \hline 
 \multicolumn{5}{|l|}{\textbf{$\varnothing$ grouped by maximal distance factor}} \\ 
1.1 & 16.74 (0.0922) & 30.85 (36.1504) & 14.1098 & 1505.9402 \\ 
  1.5 & 13.86 (0.0688) & 26.55 (27.9252) & 12.6967 & 941.0220 \\ 
  2.0 & 13.14 (0.0617) & 22.53 (21.2259) & 9.3912 & 538.7330 \\ 
   \hline 
 \multicolumn{5}{|l|}{\textbf{$\varnothing$ grouped by tuning parameter $\tau$}} \\ 
1.1 & 14.61 (0.0424) & 19.30 (14.4547) & 4.6910 & 230.9429 \\ 
  1.2 & 14.61 (0.0424) & 21.40 (16.4506) & 6.7886 & 316.7075 \\ 
  1.5 & 14.61 (0.0424) & 25.07 (22.3089) & 10.4530 & 606.9529 \\ 
  2.0 & 14.61 (0.0424) & 28.41 (29.7121) & 13.8017 & 1073.2966 \\ 
  2.5 & 14.61 (0.0424) & 30.65 (35.4179) & 16.0411 & 1511.7454 \\ 
  3.0 & 14.61 (0.0424) & 32.42 (40.2369) & 17.8072 & 1936.1004 \\ 
   \hline
\end{tabular}
\endgroup
\end{table}

% latex table generated in R 4.3.0 by xtable 1.8-4 package
% Mon Jun 12 15:50:51 2023
\begin{table}[h!]
\centering
\parbox{\textwidth}{\caption{Check Filter Method - Performance measure results for the shape effect $ \phi $ grouped by sample size, sample density, maximal distance factor and tuning parameter $\tau$.}} 
\label{perftable_check_shape}
\begingroup\footnotesize
\begin{tabular}{|L{2cm}R{3.5cm}R{3cm}R{2cm}R{2cm}|}
  \hline
semi-variogram parameter & $\eta_{\theta_i}$ ($se_{MC}(\eta_{\theta_i})$) & $\overline{\widehat{\eta_{\theta_i}}}$ ($sd(\widehat{\eta_{\theta_i}})$) & $bias(\widehat{\eta_{\theta_i}})$ & $MSE(\widehat{\eta_{\theta_i}})$ \\ 
  \hline 
 \multicolumn{5}{|l|}{\textbf{$\varnothing$ grouped by sample size}} \\ 
500 & 184.79 (1.1083) & 123.30 (171.2830) & -61.4882 & 33118.6714 \\ 
  1000 & 177.56 (0.8750) & 68.28 $\;$ (94.3200) & -109.2805 & 20838.4870 \\ 
  2000 & 144.22 (0.6455) & 43.24 $\;$ (50.1536) & -100.9771 & 12711.7636 \\ 
   \hline 
 \multicolumn{5}{|l|}{\textbf{$\varnothing$ grouped by sample density}} \\ 
low & 170.98 (0.8997) & 76.55 (119.4568) & -94.4307 & 23187.0799 \\ 
  middle & 169.24 (0.8371) & 71.42 (110.4234) & -97.8230 & 21762.6725 \\ 
  high & 158.86 (0.7759) & 64.77 $\;$ (98.4035) & -94.0835 & 18534.9506 \\ 
   \hline 
 \multicolumn{5}{|l|}{\textbf{$\varnothing$ grouped by maximal distance factor}} \\ 
1.1 & 183.22 (1.0090) & 80.61 (132.6585) & -102.6192 & 28128.9708 \\ 
  1.5 & 169.41 (0.8413) & 71.47 (105.7821) & -97.9494 & 20783.9406 \\ 
  2.0 & 149.34 (0.7010) & 62.64 $\;$ (91.7611) & -86.6964 & 15936.3747 \\ 
   \hline 
 \multicolumn{5}{|l|}{\textbf{$\varnothing$ grouped by tuning parameter $\tau$}} \\ 
1.1 & 166.43 (0.4827) & 45.00 $\;$ (63.9584) & -121.4394 & 18838.2027 \\ 
  1.2 & 166.43 (0.4827) & 52.15 $\;$ (69.1959) & -114.2826 & 17848.5779 \\ 
  1.5 & 166.43 (0.4827) & 65.55 $\;$ (85.7781) & -100.8865 & 17535.9610 \\ 
  2.0 & 166.43 (0.4827) & 78.55 (111.1835) & -87.8894 & 20086.3245 \\ 
  2.5 & 166.43 (0.4827) & 87.65 (133.4778) & -78.7840 & 24023.2575 \\ 
  3.0 & 166.43 (0.4827) & 94.91 (152.9463) & -71.5295 & 28509.0593 \\ 
   \hline
\end{tabular}
\endgroup
\end{table}

% latex table generated in R 4.3.0 by xtable 1.8-4 package
% Mon Jun 12 15:51:54 2023
\begin{table}[h!]
\centering
\parbox{\textwidth}{\caption{Quantile Filter Method - Performance measure results for the nugget effect $ c_0 $ grouped by sample size, sample density, maximal distance factor and tuning parameter $\alpha$.}} 
\label{perftable_quantile_nugget}
\begingroup\footnotesize
\begin{tabular}{|L{2cm}R{3.5cm}R{3cm}R{2cm}R{2cm}|}
  \hline
semi-variogram parameter & $\eta_{\theta_i}$ ($se_{MC}(\eta_{\theta_i})$) & $\overline{\widehat{\eta_{\theta_i}}}$ ($sd(\widehat{\eta_{\theta_i}})$) & $bias(\widehat{\eta_{\theta_i}})$ & $MSE(\widehat{\eta_{\theta_i}})$ \\ 
  \hline 
 \multicolumn{5}{|l|}{\textbf{$\varnothing$ grouped by sample size}} \\ 
500 & 19.62 (0.1177) & 6.40 (4.9385) & -13.2174 & 199.0893 \\ 
  1000 & 15.08 (0.0743) & 4.07 (2.7157) & -11.0151 & 128.7070 \\ 
  2000 & 8.58 (0.0384) & 2.59 (1.1072) & -5.9969 & 37.1887 \\ 
   \hline 
 \multicolumn{5}{|l|}{\textbf{$\varnothing$ grouped by sample density}} \\ 
low & 16.63 (0.0875) & 4.28 (3.7703) & -12.3532 & 166.8155 \\ 
  middle & 14.15 (0.0700) & 3.85 (3.0847) & -10.3045 & 115.6991 \\ 
  high & 14.01 (0.0684) & 3.89 (3.0748) & -10.1221 & 111.9110 \\ 
   \hline 
 \multicolumn{5}{|l|}{\textbf{$\varnothing$ grouped by maximal distance factor}} \\ 
1.1 & 15.58 (0.0858) & 3.92 (3.6811) & -11.6626 & 149.5658 \\ 
  1.5 & 14.71 (0.0731) & 3.93 (3.2720) & -10.7813 & 126.9426 \\ 
  2.0 & 14.36 (0.0674) & 4.1 (3.0461) & -10.2524 & 114.3899 \\ 
   \hline 
 \multicolumn{5}{|l|}{\textbf{$\varnothing$ grouped by tuning parameter $\alpha$}} \\ 
0.75 & 14.95 (0.0433) & 2.09 (1.7179) & -12.8538 & 168.1719 \\ 
  0.80 & 14.95 (0.0433) & 2.70 (2.0903) & -12.2452 & 154.3156 \\ 
  0.85 & 14.95 (0.0433) & 3.36 (2.4935) & -11.5860 & 140.4520 \\ 
  0.90 & 14.95 (0.0433) & 4.11 (2.9409) & -10.8361 & 126.0688 \\ 
  0.95 & 14.95 (0.0433) & 5.04 (3.4667) & -9.9039 & 110.1047 \\ 
  1.00 & 14.95 (0.0433) & 6.64 (4.2123) & -8.3068 & 86.7460 \\ 
   \hline
\end{tabular}
\endgroup
\end{table}

% latex table generated in R 4.3.0 by xtable 1.8-4 package
% Mon Jun 12 15:52:45 2023
\begin{table}[h!]
\centering
\parbox{\textwidth}{\caption{Quantile Filter Method - Performance measure results for the partial sill effect $ \sigma^2_0 $ grouped by sample size, sample density, maximal distance factor and tuning parameter $\alpha$.}} 
\label{perftable_quantile_partial sill}
\begingroup\footnotesize
\begin{tabular}{|L{2cm}R{3.5cm}R{3cm}R{2cm}R{2cm}|}
  \hline
semi-variogram parameter & $\eta_{\theta_i}$ ($se_{MC}(\eta_{\theta_i})$) & $\overline{\widehat{\eta_{\theta_i}}}$ ($sd(\widehat{\eta_{\theta_i}})$) & $bias(\widehat{\eta_{\theta_i}})$ & $MSE(\widehat{\eta_{\theta_i}})$ \\ 
  \hline 
 \multicolumn{5}{|l|}{\textbf{$\varnothing$ grouped by sample size}} \\ 
500 & 19.68 (0.1180) & 558.20 (25208.56) & 538.52 & $3.36\cdot 10^8$ \\ 
  1000 & 13.76 (0.0678) & 355.86 (28324.53) & 342.10 & $8.02\cdot 10^8$\\ 
  2000 & 8.63 (0.0386) & 77.29 (13782.64) & 68.66 & $1.90\cdot 10^8$ \\ 
   \hline 
 \multicolumn{5}{|l|}{\textbf{$\varnothing$ grouped by sample density}} \\ 
low & 16.24 (0.0855) & 350.68 (27205.76) & 334.43 & $7.40\cdot 10^8$ \\ 
  middle & 13.84 (0.0685) & 251.10 (17413.02) & 237.26 & $3.03\cdot 10^8$ \\ 
  high & 13.7 (0.0669) & 264.91 (22420.21) & 251.21 & $5.03\cdot 10^8$ \\ 
   \hline 
 \multicolumn{5}{|l|}{\textbf{$\varnothing$ grouped by maximal distance factor}} \\ 
1.1 & 16.74 (0.0922) & 564.73 (30973.09) & 547.99 & $9.60\cdot 10^8$ \\ 
  1.5 & 13.86 (0.0688) & 173.29 (13364.20) & 159.43 & $1.79\cdot 10^8$ \\ 
  2.0 & 13.14 (0.0617) & 184.74 (21672.27) & 171.59 & $4.70\cdot 10^8$ \\ 
   \hline 
 \multicolumn{5}{|l|}{\textbf{$\varnothing$ grouped by tuning parameter $\alpha$}} \\ 
0.75 & 14.61 (0.0424) & 17.72$\;\;\;\;\;\;\;$(16.86) & 3.11 & 293.89 \\ 
  0.80 & 14.61 (0.0424) & 19.18$\;\;\;\;\;\;\;$(19.60) & 4.57 & 404.83 \\ 
  0.85 & 14.61 (0.0424) & 21.03$\;\;\;\;\;\;\;$(23.48) & 6.42 & 592.40 \\ 
  0.90 & 14.61 (0.0424) & 23.69$\;\;\;\;\;\;\;$(29.86) & 9.08 & 974.26 \\ 
  0.95 & 14.61 (0.0424) & 28.65$\;\;\;\;\;\;\;$(44.45) & 14.03 & 2172.62 \\ 
  1.00 & 14.61 (0.0424) & 1606.99 (55093.93) & 1592.38 & $3.04\cdot 10^9$ \\ 
   \hline
\end{tabular}
\endgroup
\end{table}

% latex table generated in R 4.3.0 by xtable 1.8-4 package
% Mon Jun 12 15:53:30 2023
\begin{table}[h!]
\centering
\parbox{\textwidth}{\caption{Quantile Filter Method - Performance measure results for the shape effect $ \phi $ grouped by sample size, sample density, maximal distance factor and tuning parameter $\alpha$.}} 
\label{perftable_quantile_shape}
\begingroup\footnotesize
\begin{tabular}{|L{2cm}R{3.5cm}R{3cm}R{2cm}R{2cm}|}
  \hline
semi-variogram parameter & $\eta_{\theta_i}$ ($se_{MC}(\eta_{\theta_i})$) & $\overline{\widehat{\eta_{\theta_i}}}$ ($sd(\widehat{\eta_{\theta_i}})$) & $bias(\widehat{\eta_{\theta_i}})$ & $MSE(\widehat{\eta_{\theta_i}})$ \\ 
  \hline 
 \multicolumn{5}{|l|}{\textbf{$\varnothing$ grouped by sample size}} \\ 
500 & 184.79 (1.1083) & 2467.08 (100233.56) & 2282.29 & $1.01\cdot10^{10}$ \\ 
  1000 & 177.56 (0.875) & 1173.72$\;\;\;$(92747.11) & 996.17 & $8.60\cdot10^9$ \\ 
  2000 & 144.22 (0.6455) & 157.78$\;\;\;$(28836.16) & 13.56 & $8.32\cdot 10^8$ \\ 
   \hline 
 \multicolumn{5}{|l|}{\textbf{$\varnothing$ grouped by sample density}} \\ 
low & 170.98 (0.8997) & 1356.23$\;\;\;$(89648.65) & 1185.25 & $8.04\cdot10^9$ \\ 
  middle & 169.24 (0.8371) & 859.64$\;\;\;$(57975.26) & 690.41 & $3.36\cdot10^9$ \\ 
  high & 158.86 (0.7759) & 970.59$\;\;\;$(76846.18) & 811.74 & $5.91\cdot10^9$ \\ 
   \hline 
 \multicolumn{5}{|l|}{\textbf{$\varnothing$ grouped by maximal distance factor}} \\ 
1.1 & 183.22 (1.009) & 1858.85$\;\;\;$(86269.15) & 1675.63 & $7.45\cdot10^9$ \\ 
  1.5 & 169.41 (0.8413) & 641.66$\;\;\;$(48402.07) & 472.25 & $2.34\cdot10^9$ \\ 
  2.0 & 149.34 (0.701) & 826.04$\;\;\;$(85884.05) & 676.71 & $7.38\cdot10^9$ \\ 
   \hline 
 \multicolumn{5}{|l|}{\textbf{$\varnothing$ grouped by tuning parameter $\alpha$}} \\ 
0.75 & 166.43 (0.4827) & 38.21$\;\;\;\;\;\;\;\;\;$(55.50) & -128.23 & 19523.01 \\ 
  0.80 & 166.43 (0.4827) & 42.65$\;\;\;\;\;\;\;\;\;$(66.53) & -123.79 & 19749.24 \\ 
  0.85 & 166.43 (0.4827) & 48.53$\;\;\;\;\;\;\;\;\;$(82.68) & -117.91 & 20738.49 \\ 
  0.90 & 166.43 (0.4827) & 57.37$\;\;\;\;\;\;\;$(109.99) & -109.07 & 23993.38 \\ 
  0.95 & 166.43 (0.4827) & 74.97$\;\;\;\;\;\;\;$(175.01) & -91.46 & 38990.93 \\ 
  1.00 & 166.43 (0.4827) & 6035.77 (184498.18) & 5869.34 & $3.41\cdot10^{10}$ \\ 
   \hline
\end{tabular}
\endgroup
\end{table}

\end{appendices}

\end{document}